\documentclass[3p,times,procedia]{elsarticle}
\usepackage{nupha_ecrc}

\volume{00}

\firstpage{1}

\journalname{Nuclear Physics A}

\runauth{}

\jid{nupha}

\jnltitlelogo{Nuclear Physics A}

\usepackage{amssymb}

\biboptions{square,comma,numbers,sort&compress}

\usepackage[figuresright]{rotating}

\usepackage{graphicx}
\usepackage{dcolumn}
\usepackage{bm}
\usepackage{amssymb}
\usepackage{xspace}

\usepackage{hyperref}
\usepackage{amsmath}

\usepackage{color}

\newcommand{\pp}{\ensuremath{\rm pp}\xspace}
\newcommand{\pA}{\ensuremath{\rm pA}\xspace}

\newcommand{\pt}{\ensuremath{p_{\rm{T}}}\xspace}

\newcommand{\mpt}{\ensuremath{\langle p_{\rm T} \rangle \xspace}}

\newcommand{\sppt}[1]{\ensuremath{\sqrt{s} = #1 \text{\,TeV}}\xspace}
\newcommand{\so}{\ensuremath{S_{\rm 0}}\xspace}
\newcommand{\nch}{\ensuremath{N_{\rm ch}}\xspace}
\newcommand{\nmpi}{\ensuremath{N_{\rm MPI}}\xspace}

\begin{document}

\begin{frontmatter}



\dochead{}

\title{Effects produced by multi-parton interactions and color reconnection in small systems}


\author{Eleazar Cuautle, Antonio Ortiz and Guy Pai{\'c}}

\address{Instituto de Ciencias Nucleares, Universidad Nacional Aut\'onoma de M\'exico. \\ Circuito exterior s/n, Ciudad Universitaria, Del. Coyoac\'an, C.P. 04510, M\'exico D. F.}

\begin{abstract}

Multi-parton interactions and color reconnection can produce QGP-like effects in small systems, specifically, radial flow-like patterns. For pp collisions simulated with Pythia 8.212, in this work we investigate their effects on different observables like event multiplicity, event shapes and transverse momentum distributions.
\end{abstract}


\end{frontmatter}


\section{Introduction}
\label{sec:1}

The \pp collisions at the LHC raise a very important question: are the pp collisions the proper ``benchmark'' for comparison with heavy-ion results? From the theoretical point of view, due to the small overlapping transverse area the initial energy densities achieved in different systems are very similar and much superior to that required for the QCD phase transition. From the experimental point of view,  several signatures considered proofs of collective effects in heavy-ion collisions have been observed in the small systems created in \pp and \pA collisions~\cite{Adam:2016dau}. Such a development has important consequences for the analysis of AA data~\cite{Zakharov:2014mda,Abelev:2014laa,Adam:2015kca} because one should take into account the contribution of the QGP-like effects in \pp when comparing with AA.  
Recently, new ideas have been investigated to understand the effects in high multiplicity \pp collisions. For example, the possibility of producing collective-like effects with multi-parton interactions (MPI) and color reconnection (CR) through boosted color strings  in QCD MC generators like Pythia~\cite{Ortiz:2013yxa} and DYPSI~\cite{Bierlich:2014xba}.  Presently, the rise of the mean transverse momentum (\mpt)  with multiplicity (\nch) and  the mass of the produced particles in QCD inspired generators are taken as proofs of the strong interaction among the color strings~\cite{Sjostrand:2013cya,Ortiz:2015cma}. Using Pythia 8.2~\cite{Sjostrand:2014zea}, in the present work we report the interrelations among different quantities like the number of MPI, the event structure (spherocity), the measured pseudorapidity range, the charged particle density  (${\rm d}N_{\rm ch}/{\rm d}\eta$) on \mpt\,and the \pt spectra. The simulations were made for \pp collisions at \sppt{7}. 
%
%

\begin{figure}
\begin{center}
\resizebox{0.95\textwidth}{!}{%
  \includegraphics{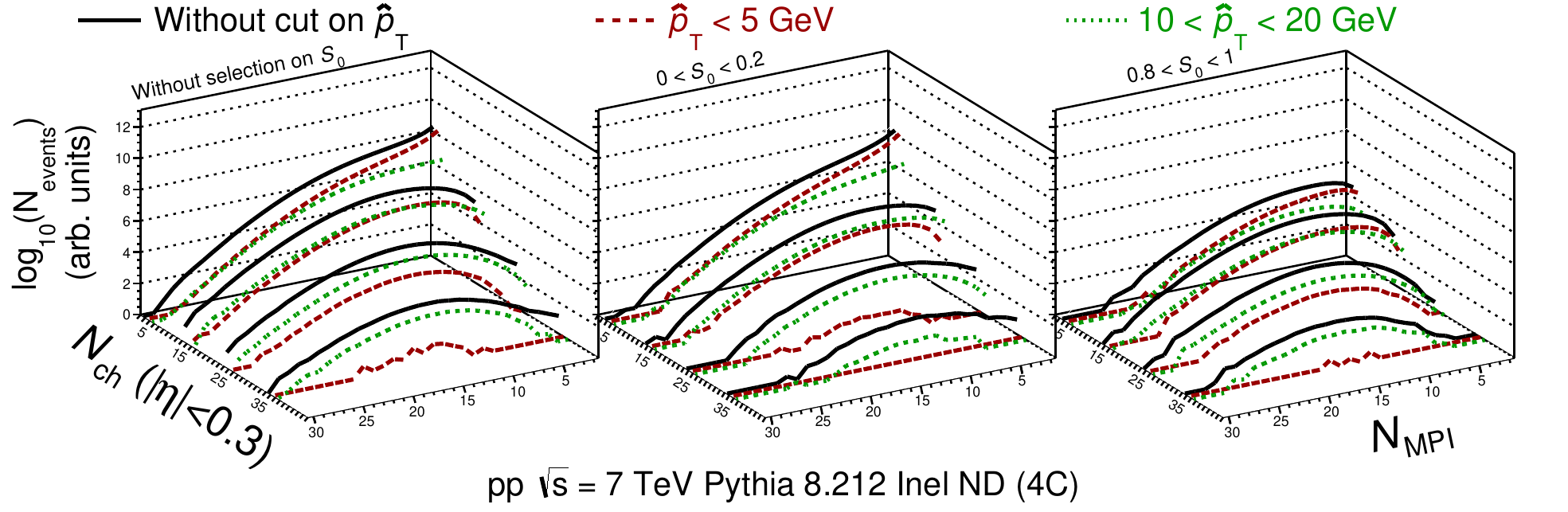}
}
\caption{(Color online). Distributions of number of multi-parton interactions as a function of the event multiplicity and leading partonic \pt ($\hat{p}_{\rm T}$) for non-diffractive (ND) inelastic \pp collisions at \sppt{7}. Three cases are displayed, from left to right, inclusive, jetty-like events (low spherocity) and isotropic events (high spherocity). }
\label{fig:1}       
\end{center}
\end{figure}

\section{Results}

To study and identify different kind  of events we use the mid-rapidity charged hadron transverse spherocity, \so. The restriction to the transverse plane avoids  the  bias   from  the  boost   along  the
beam axis~\cite{Banfi:2004nk}. It is defined for an unit transverse vector $\mathbf{\hat{n}}$ which minimizes the ratio below:

\begin{equation}
S_{\rm 0} = \frac{\pi^{2}}{4} \left(  \frac{\sum_{i} |{{\overrightarrow{\pt}}}_{i} \times \mathbf{\hat{n}}|}{\sum_{i} {\pt}_{i}}  \right)^{2}
\end{equation}

\begin{figure}
\begin{center}
\resizebox{0.45\textwidth}{!}{%
  \includegraphics{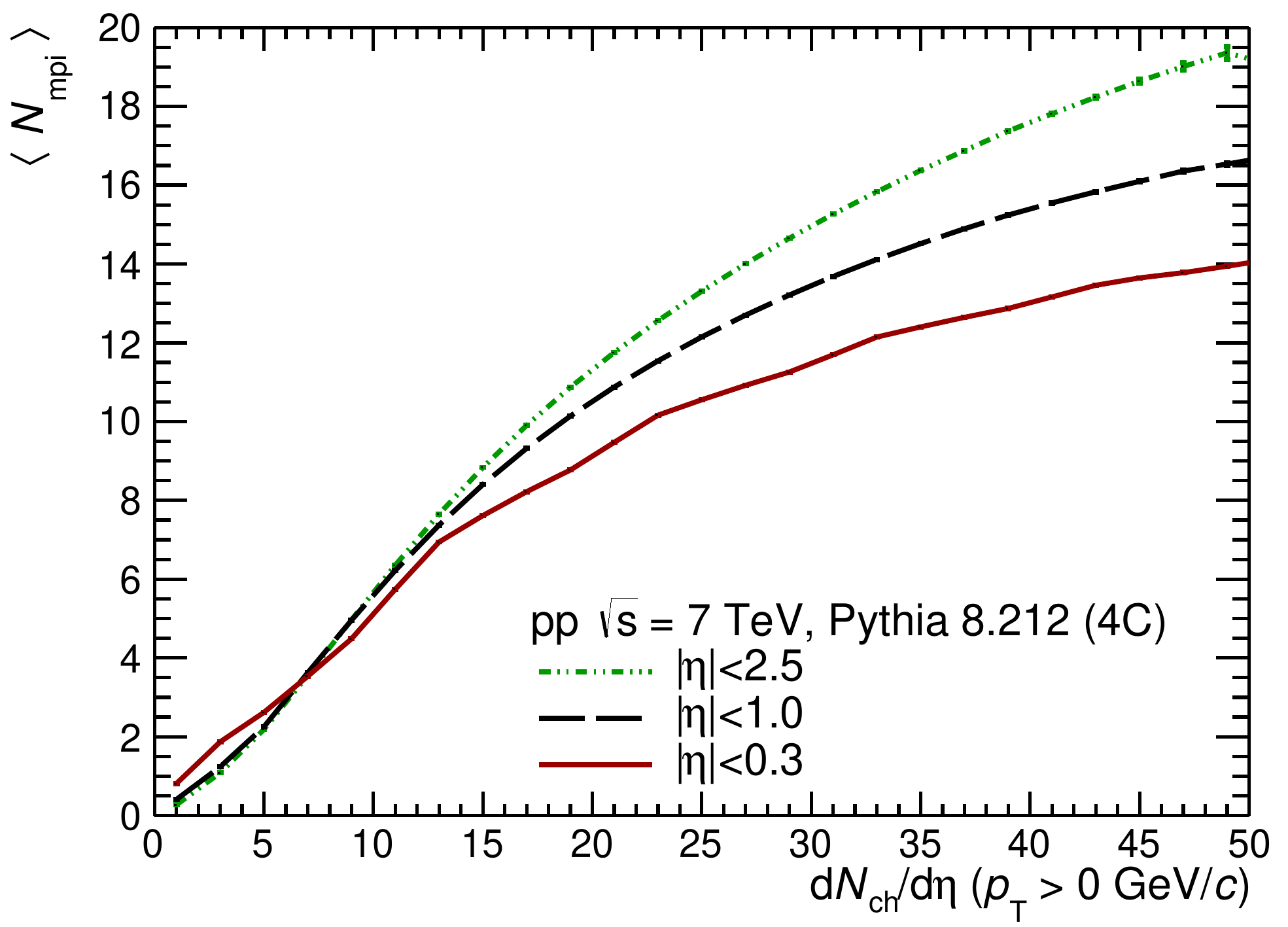}
}
\caption{(Color online). Average number of multi-parton interactions as a function of the charged particle density (${\rm d}N_{\rm ch}/{\rm d}\eta$) for \pp collisions at \sppt{7}. The charged particle density is calculated for three pseudorapidity intervals.}
\label{fig:2}       
\end{center}
\end{figure}

By construction, spherocity close to 0 (1) is related to events which are jetty-like (isotropic) in the transverse plane. To calculate the transverse spherocity and multiplicity, we consider only charged hadrons with momenta above 0.15\,GeV/$c$~\cite{Cuautle:2014yda} and $|\eta|$\,$<$\,$0.3$.  Figure~\ref{fig:1} shows  the distribution of number of MPI (\nmpi) as a function of multiplicity for non-diffractive inelastic \pp collisions and for two extreme spherocity bins, $S_{\rm 0}$\,$<$\,$0.2$ (jetty-like) and $S_{\rm 0}$\,$>$\,$0.8$ (isotropic).  Also shown are the results for two intervals of the leading parton transverse momentum ($\hat{p}_{\rm T}$) of the event: $\hat{p}_{\rm T}$\,$<$\,$5$\,GeV and $10$\,$<$\,$\hat{p}_{\rm T}$\,$<$\,$20$\,GeV. We observe that the average number of multi parton interactions increases with increasing multiplicity, slightly more so larger for isotropic events. At low multiplicity the dominant events are those of low \nmpi and with $\hat{p}_{\rm T}$\,$<$\,$5$\,GeV. On the other hand, in high multiplicity events the number of events with $10$\,$<$\,$\hat{p}_{\rm T}$\,$<$\,$20$\,GeV is greater than those with $\hat{p}_{\rm T}$\,$<$\,$5$\,GeV. In Pythia, high multiplicity events are therefore produced via hard partonic scatterings as discussed in~\cite{Abelev:2012sk}, where data indicate that soft processes dominate.  It is worth noticing that the event is isotropic or jetty-like within the restricted $\eta$-range which is considered for the calculation of the spherocity. This suggests that when selecting isotropic events we study only the soft component of the \pp interactions, i.e., the underlying activity accompanying the high \pt jets, while jetty-like events are those which have poor underlying event activity within the small acceptance under consideration.


To investigate the acceptance effects we studied some observables calculated at mid-rapidity ($|\eta|$\,$<$\,$0.3$), as a function of the charged particle density obtained for different pseudorapidity intervals, namely, $|\eta|$\,$<$\,$0.3$, $|\eta|$\,$<$\,$1.0$ and $|\eta|$\,$<$\,$2.5$. Figure~\ref{fig:2} shows the average \nmpi as a function of the charged particle density. For ${\rm d}N_{\rm ch}/{\rm d}\eta$\,$<$\,$5$, the average \nmpi shows little or no dependence with the pseudorapidity interval used to calculate ${\rm d}N_{\rm ch}/{\rm d}\eta$.  While for ${\rm d}N_{\rm ch}/{\rm d}\eta$\,$>$\,$5$, the average \nmpi exhibits the opposite behavior, for instance, at ${\rm d}N_{\rm ch}/{\rm d}\eta$\,$=$\,$30$, the average \nmpi increases by $\approx$30\% going from $|\eta|$\,$<$\,$0.3$ to $|\eta|$\,$<$\,$2.5$.

\begin{figure}[htbp]
\begin{center}
   \includegraphics[width=0.45\textwidth]{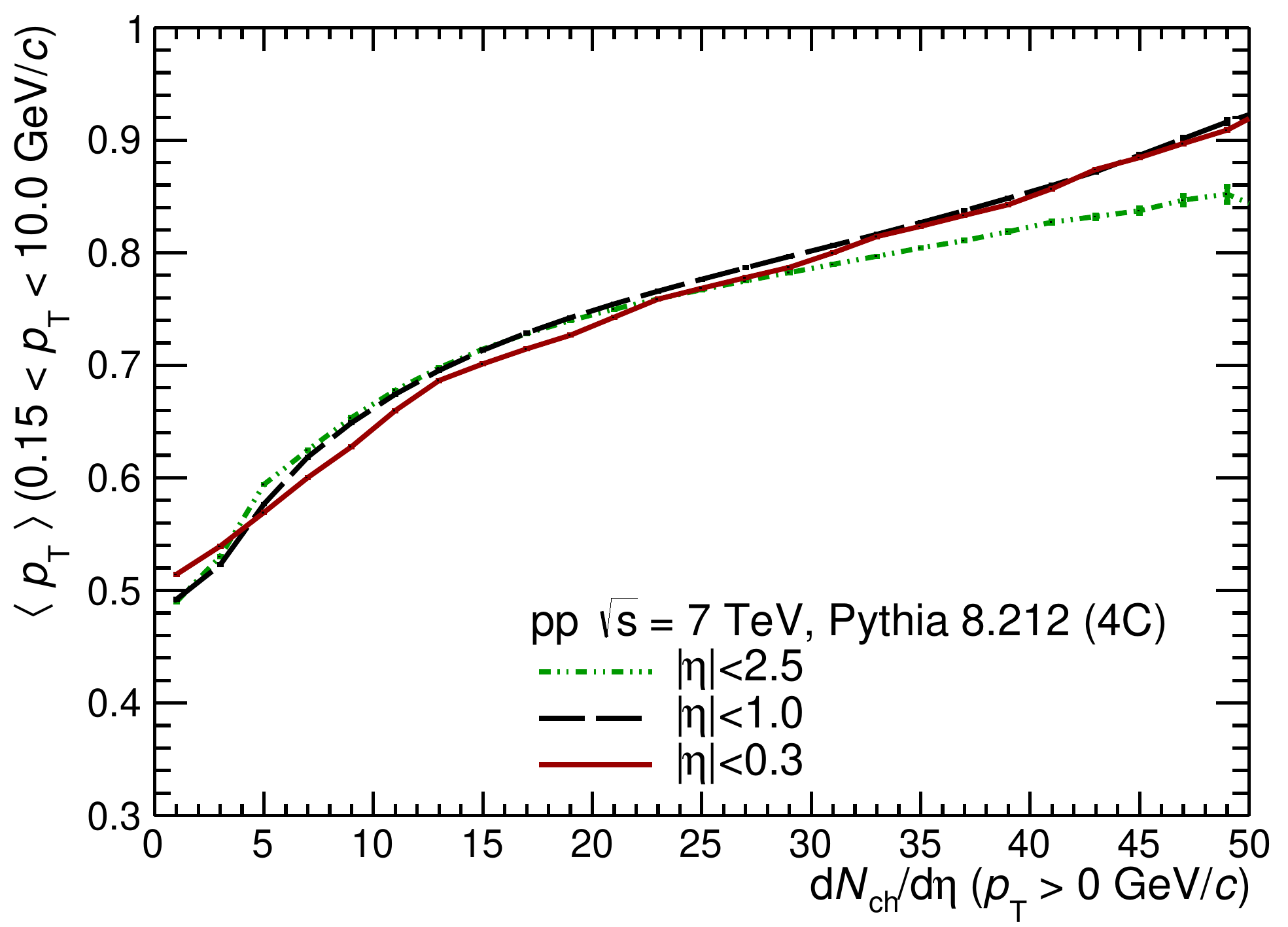}   
   \includegraphics[width=0.45\textwidth]{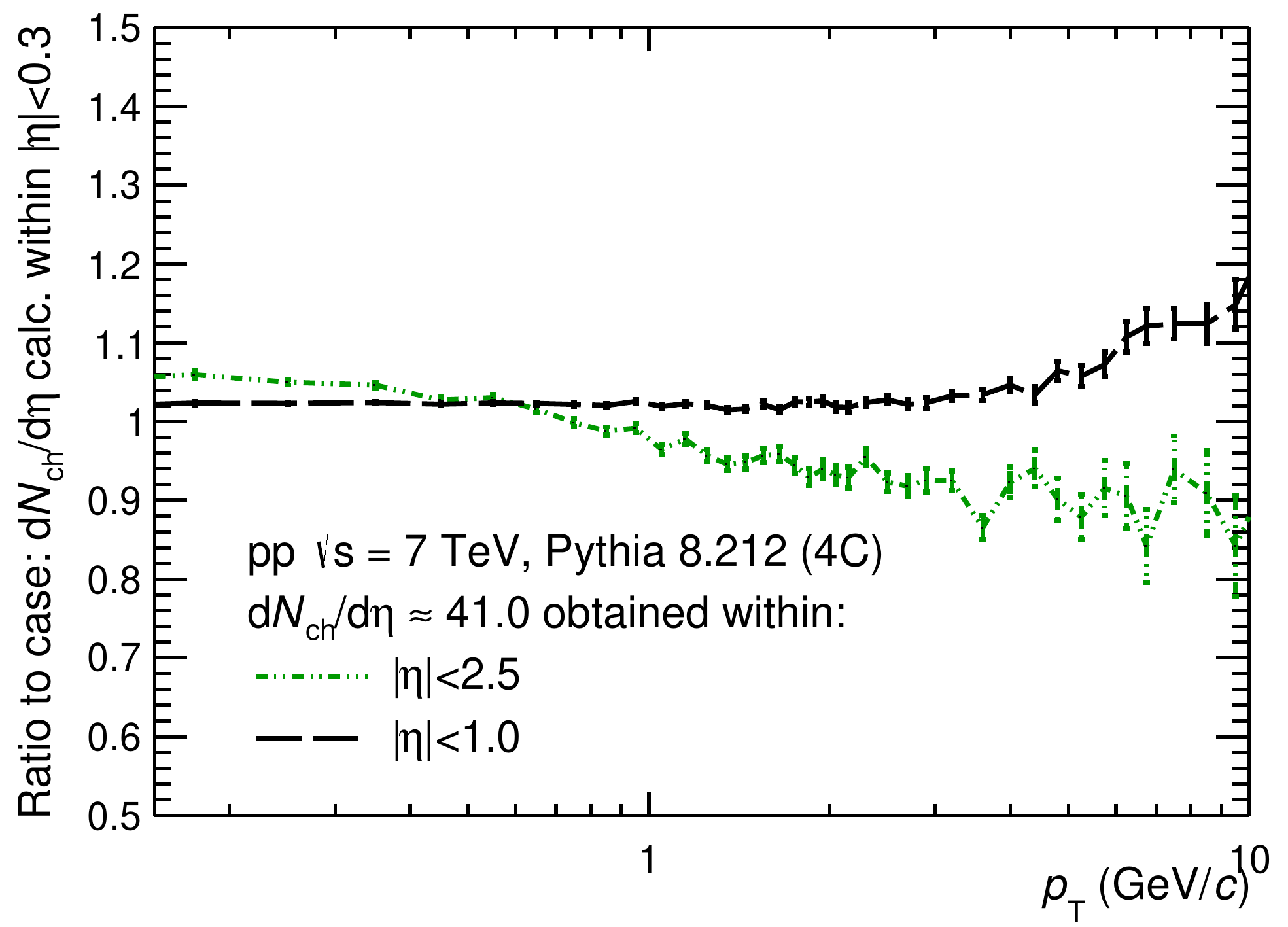}
   \caption{(Color online).  Average transverse momentum as a function of the charged particle density (${\rm d}N_{\rm ch}/{\rm d}\eta$) for \pp collisions at \sppt{7}, where, ${\rm d}N_{\rm ch}/{\rm d}\eta$ has been computed for different pseudorapidity intervals (left). Transverse momentum spectra, obtained for  ${\rm d}N_{\rm ch}/{\rm d}\eta$\,$\approx$\,$41$ calculated within $|\eta|$\,$<$\,$1$ and $|\eta|$\,$<$\,$2.5$, have been normalised to that for $|\eta|$\,$<$\,$0.3$ (right). }
  \label{fig:3}
\end{center}
\end{figure}

We see in the results of ALICE reported in~\cite{Abelev:2013bla} a suggestion of a second rise of \mpt\, vs. $N_{\rm ch}$ at ${\rm d}N_{\rm ch}/{\rm d}\eta$\,$\approx$\,$41$ for \pp collisions at \sppt{7}~\cite{Abelev:2013bla}, the effect can be understood as due to the multiplicity selection which biases the sample towards hard events. To study the impact of this bias on global observables, Fig.~\ref{fig:3} shows the average transverse momentum as a function of ${\rm d}N_{\rm ch}/{\rm d}\eta$.  When the same kinematic cuts like those used in~\cite{Abelev:2013bla} are implemented (multiplicity and \mpt\, calculated within $|\eta|$\,$<$\,$0.3$), we observe a second rise. However, the correlation between the event activity and the hardness of the event disappears calculating the multiplicity within a wider $|\eta|$ range. The results suggest that it may be important to carefully consider the pseudorapidity range used in the measurement because according with the simulations the events are differing in their hardness. For ${\rm d}N_{\rm ch}/{\rm d}\eta$\,$\approx$\,$41$, the \mpt's are the same for the cases where ${\rm d}N_{\rm ch}/{\rm d}\eta$ is calculated within $|\eta|$\,$<$\,$1$ and $|\eta|$\,$<$\,$0.3$. Actually, the spectral shapes are the same for $\pt$\,$<$\,$5$\,GeV/$c$ as shown in the right hand side plot of Fig.~\ref{fig:3}. On the other hand, the case $|\eta|$\,$<$\,$2.5$ gives a \pt spectrum which is softer (effect of order $\approx$5\%) than those for $|\eta|$\,$<$\,$1$ and $|\eta|$\,$<$\,$0.3$.

\begin{figure}[htbp]
\begin{center}
   \includegraphics[width=0.45\textwidth]{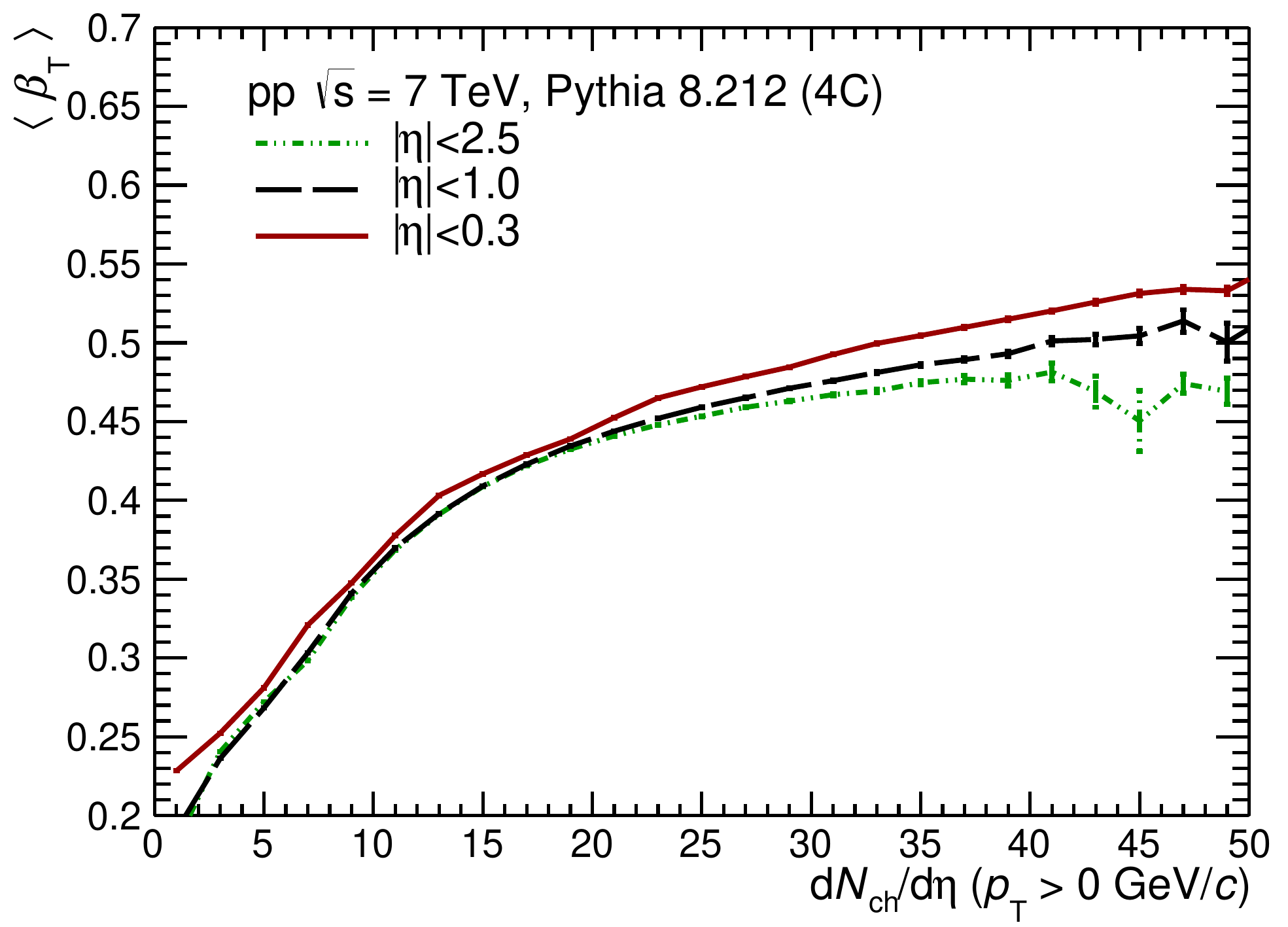}   
   \includegraphics[width=0.45\textwidth]{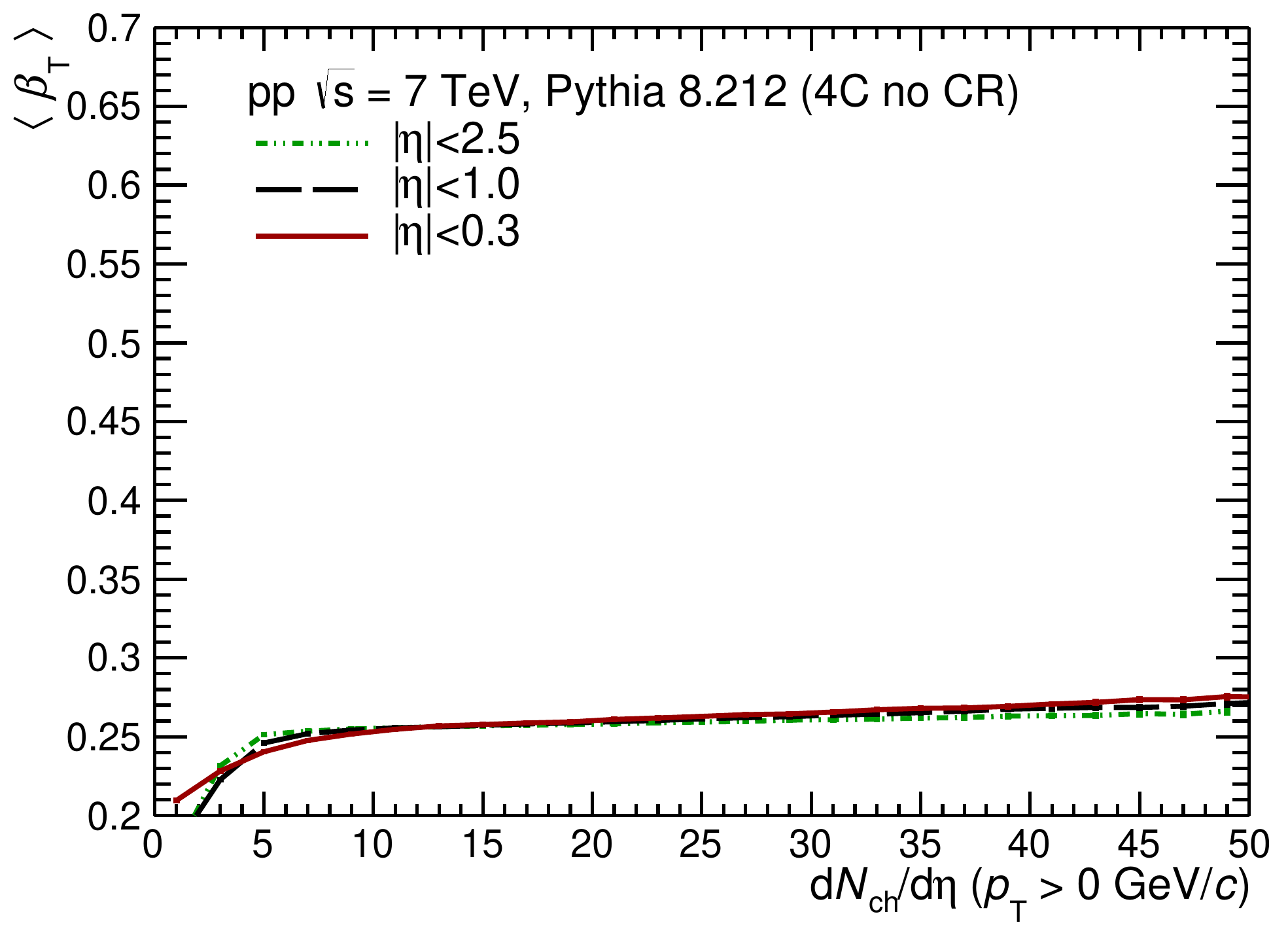}
   \caption{(Color online).  Average transverse velocity extracted from the blast-wave analysis as a function of the charged particle density (${\rm d}N_{\rm ch}/{\rm d}\eta$) for \pp collisions at \sppt{7}. The charged particle density has been computed for different pseudorapidity intervals. Results for cases with (left) and without (right) color reconnection are displayed.}
  \label{fig:4}
\end{center}
\end{figure}

The effect of the jet bias has been also investigated using identified hadrons. For this a simultaneous fit of the blast-wave function to the \pt spectra of the particles listed in~\cite{Cuautle:2015kra} has been implemented. The parameters extracted from the fits are studied as a function of event multiplicity. Transverse momentum distributions and transverse spherocity are obtained using only primary charged hadrons with transverse momenta above 0.15 GeV/$c$ and $|\eta|$\,$<$\,$0.3$. The event multiplicity is calculated counting primary charged hadrons inside different pseudorapidity intervals. Figure~\ref{fig:4} shows the average transverse velocity as a function of the charged particle density. When the jet bias is the strongest, i.e., when the event multiplicity and \pt spectra are obtained within the same pseudorapidity interval ($|\eta|$\,$<$\,$0.3$), the average transverse velocity exhibits always a rise with the multiplicity. On the other hand, when the correlation is weakened e.g., when multiplicity is calculated in a wider pseudorapidity interval, the average transverse expansion velocity is smaller and seems to reach a saturation for ${\rm d}N_{\rm ch}/{\rm d}\eta$\,$>$\,$30$. This result agrees with that reported earlier~\cite{Cuautle:2015kra}, where it was shown that jets increase the apparent radial expansion velocity extracted from the blast-wave analysis. In that study, the event classification was done using transverse spherocity. The right hand side plot of Fig.~\ref{fig:4} shows that even without CR the jet bias effect is also visible, i.e., $\langle  \beta_{\rm T} \rangle$ is slightly higher when the charged particle density is calculated in the narrowest $\eta$ interval. The effects discussed here may play a role in the measurements reported by experiments at the LHC, where the events are classified according with their multiplicity measured within different pseudorapidity intervals.

\section{Conclusions}

The present study brings an insight in the correlation between the number of multi-parton interactions and the hardness/multiplicity of the event. Most importantly, we demonstrate the importance of the color reconnection for the final state interaction in \pp collisions which brings collective effects without invoking classical hydrodynamical arguments. Finally, we show that different observables studied with Pythia are sensitive to the pseudorapidity interval were the charged particle density is measured, suggesting that reporting measurements of variables in the central pseudorapidity/rapidity region for events whose multiplicity is obtained in a different pseudorapidity range may not be bias free.

Support for this work has been received from CONACYT under the grant No. 260440; and from DGAPA-UNAM under PAPIIT grants IA102515, IN105113, IN107911 and IN108414.

\bibliographystyle{elsarticle-num}

\bibliography{biblio}

\begin{thebibliography}{10}
\expandafter\ifx\csname url\endcsname\relax
  \def\url#1{\texttt{#1}}\fi
\expandafter\ifx\csname urlprefix\endcsname\relax\def\urlprefix{URL }\fi
\expandafter\ifx\csname href\endcsname\relax
  \def\href#1#2{#2} \def\path#1{#1}\fi

\bibitem{Adam:2016dau}
J.~Adam, et~al., {Multiplicity dependence of charged pion, kaon, and
  (anti)proton production at large transverse momentum in p-Pb collisions at
  $\mathbf{\sqrt{{\textit s}_{\rm NN}}}$ = 5.02 TeV, }\href
  {http://arxiv.org/abs/1601.03658} {\path{arXiv:1601.03658}}.

\bibitem{Zakharov:2014mda}
B.~G. Zakharov, {Jet quenching in pp and pA collisions}, in: {11th Conference
  on Quark Confinement and the Hadron Spectrum (Confinement XI) St. Petersburg,
  Russia, September 8-12, 2014}, 2014.
\newblock \href {http://arxiv.org/abs/1412.0295} {\path{arXiv:1412.0295}}.

\bibitem{Abelev:2014laa}
B.~B. Abelev, et~al., Phys. Lett. B736 (2014) 196--207.

\bibitem{Adam:2015kca}
J.~Adam, et~al., {Centrality dependence of the nuclear modification factor of
  charged pions, kaons, and protons in Pb-Pb collisions at $\sqrt{s_{\rm
  NN}}=2.76$ TeV, }\href {http://arxiv.org/abs/1506.07287}
  {\path{arXiv:1506.07287}}.

\bibitem{Ortiz:2013yxa}
A.~Ortiz, P.~Christiansen, E.~Cuautle, I.~Maldonado, G.~Pai{\'c}, Phys. Rev.
  Lett. 111~({\bf 4}) (2013) 042001.

\bibitem{Bierlich:2014xba}
C.~Bierlich, G.~Gustafson, L.~L{\"o}nnblad, A.~Tarasov, JHEP 03 (2015) 148.

\bibitem{Sjostrand:2013cya}
{Sj{\"o}strand, Torbj{\"o}rn}, {Colour reconnection and its effects on precise
  measurements at the LHC}, in: {Proceedings, 48th Rencontres de Moriond on QCD
  and High Energy Interactions, }, 2013, pp. 247--251.
\newblock \href {http://arxiv.org/abs/1310.8073} {\path{arXiv:1310.8073}}.

\bibitem{Ortiz:2015cma}
A.~Ortiz, Nucl. Phys. A943 (2015) 9--17.

\bibitem{Sjostrand:2014zea}
T.~Sj{\"o}strand, S.~Ask, J.~R. Christiansen, R.~Corke, N.~Desai, P.~Ilten,
  S.~Mrenna, S.~Prestel, C.~O. Rasmussen, P.~Z. Skands, Comput. Phys. Commun.
  191 (2015) 159--177.

\bibitem{Banfi:2004nk}
A.~Banfi, G.~P. Salam, G.~Zanderighi, {Resummed event shapes at hadron - hadron
  colliders}, JHEP 08 (2004) 062.

\bibitem{Cuautle:2014yda}
E.~Cuautle, R.~Jimenez, I.~Maldonado, A.~Ortiz, G.~Pai{\'c}, E.~Perez,
  {Disentangling the soft and hard components of the pp collisions using the
  sphero(i)city approach, }\href {http://arxiv.org/abs/1404.2372}
  {\path{arXiv:1404.2372}}.

\bibitem{Abelev:2012sk}
B.~Abelev, et~al., Eur. Phys. J. C72 (2012) 2124.

\bibitem{Abelev:2013bla}
B.~B. Abelev, et~al., Phys. Lett. B727 (2013) 371--380.

\bibitem{Cuautle:2015kra}
A.~Ortiz, G.~Pai{\'c}, E.~Cuautle, Nucl. Phys. A941 (2015) 78--86.

\end{thebibliography}

\end{document}